# Semiconducting behaviors at epitaxial $Ca_{0.5}TaO_3$ interfaces

Guangdong Nie[1, 2, #], Guanghui Han[3, #], Shengpu Huang[4], Huiyin Wu[1, 2], Deshun Wang[1, 2], Kangxi Liu[1], Hao Ding[1, 2], Fangdong Tang[1, 2], Licong Peng[3], Dashuai Ma[2, 4], Young Sun[1, 2], Changjiang Liu[5] & Deshun Hong[1,2*]

[1]Department of Applied Physics, Chongqing University, Chongqing 401331, China

[2]Center of Quantum Materials and Devices, Chongqing University, Chongqing 401331, China

[3]School of Materials Science and Engineering, Peking University, Beijing 100871, China

[4]Institute for Structure and Function, Department of Physics, and Chongqing Key Laboratory for Strongly Coupled Physics, Chongqing University, Chongqing 400044, China

[5]Department of Physics, University at Buffalo, SUNY, Buffalo, NY, 14260, USA

*To whom correspondence should be addressed: dhong@cqu.edu.cn

**Abstract.** Emergent phenomena take place in symmetry-breaking systems, notably the recently discovered two-dimensional electron gas and its tunable superconductivities near the $KTaO_3$ interfaces. Here, we synthesized perovskite $Ca_{0.5}TaO_3$ films along both [001] and [111] orientations. Different from the $KTaO_3$ system, $Ca_{0.5}TaO_3$ films show semiconducting behaviors when capped with $LaAlO_3$ films in both [001] and [111] orientations. By growing films at higher temperatures, more oxygen vacancies can be introduced, and the carrier density can be tuned from ~ $10^{14}$ $cm^{-2}$ to ~ $10^{16}$ $cm^{-2}$. Another difference is that the superconducting transition temperature $T_c$ in $KTaO_3$ (111) increases linearly along with its carrier density, while the $Ca_{0.5}TaO_3$ (111) remains semiconducting when carrier density ranges from ~ $10^{14}$ $cm^{-2}$ to ~ $10^{16}$ $cm^{-2}$. Based on the density function theory calculation, $Ca_{0.5}TaO_3$ and $KTaO_3$ show similar electronic band structures. According to the energy-dispersive X-ray spectroscopy, we found heavy Sr diffusion from the substrate to the $Ca_{0.5}TaO_3$ layer, which may destroy the interfacial conductivity. Our work demonstrates that besides the oxygen vacancies, electronic transport is sensitive to the atomic intermixing near the interface in tantulates.

"The interface is the device" [1], firstly referring to the semiconductor films for photonic and electronic applications half a century ago, became more applicable after the discovery of the interfacial two-dimensional electron gas in $LaAlO_3/SrTiO_3$ (LAO/STO) in 2004 [2]. In LAO/STO, high mobility conduction [2, 3], large spin-to-charge conversion efficiency [4, 5], together with the superconducting transition at 200 mK [6] have long been the research spotlight for both fundamental research and device application points of view. Recently, the KTaO$_3$ interface was found to be conductive and superconducting when capped with oxide insulators, such as EuO and LaAlO$_3$ [7, 8]. Although KTaO$_3$ shares similarities with SrTiO$_3$ (such as similar crystal structure, band structure, and quantum paraelectricity), the interfacial properties vary. In KTaO$_3$, the superconductivity develops at $T_c \sim 2$ K in the (111) plane [7, 8], and $T_c \sim 0.9$ K in the (110) plane [9, 10, 11, 12], which is about one order higher than the $T_c$ in LAO/STO. Besides, the superconductivity in KTaO$_3$ is orientation selective, especially in the low-carrier density regime [11, 12]. Until now, the origin of these different properties in superconductivity between SrTiO$_3$ and KTaO$_3$ is still under debate, and possible explanations such as larger spin-orbit coupling in KTaO$_3$, and inter-orbital and longitudinal optical phonons mediated pairing mechanisms are proposed [11, 13, 14, 15].

As the second family of interfacial superconducting perovskites, searching for more tantalate perovskites is essential for addressing the mechanism of their peculiar physical properties. One straightforward way is replacing the A site potassium in KTaO$_3$ with other alkali or alkaline-earth elements. LiTaO$_3$, a commercially available ferroelectric single crystal, happens to be rhombohedral [16]. For the highly reactive elements Rb and Cs, the RbTaO$_3$ and CsTaO$_3$ tend to be monoclinic and orthorhombic respectively [17, 18]. Notably, NaTaO$_3$ and Ca$_{0.5}$TaO$_3$ can crystalize in perovskite structures with lattice constant close to KTaO$_3$ (3.875 Å, 3.931 Å, and 3.989 Å for Ca$_{0.5}$TaO$_3$, NaTaO$_3$ and KTaO$_3$ respectively) [19, 20, 21]. Considering the volatility, Ca$_{0.5}$TaO$_3$ is easier to synthesize than NaTaO$_3$. Here, we grew [001] and [111] oriented Ca$_{0.5}$TaO$_3$ films on (LaSr)(AlTa)O$_3$ (LSAT) substrates and systematically studied their interfacial transport behaviors. Different from KTaO$_3$, our results show that the Ca$_{0.5}$TaO$_3$ interfaces are semiconducting, despite the large carrier density observed in these samples. Our DFT calculations suggest that Ca$_{0.5}$TaO$_3$ and KTaO$_3$ share similar electronic structures. However, the non-superconducting behavior at the Ca$_{0.5}$TaO$_3$ interfaces indicates the critical role of specific phonon modes in KTaO$_3$. Additionally, our EDS element mapping reveals significant intermixing of Sr with Ca$_{0.5}$TaO$_3$, which might be a primary factor in suppressing the metallicity of these interfaces.

In perovskite Ca$_{0.5}$TaO$_3$, 50% of A sites are unoccupied, as shown in Fig. 1 (a). Our Ca$_{0.5}$TaO$_3$ samples were synthesized by pulsed laser deposition (see Supplemental Materials [22]). To reduce the lattice mismatch, LSAT (001) and (111) single-crystal substrates were used during the growth. The optimized growth parameters are 800 °C and 1 Pa $O_2$ for both orientations. After growth, the samples were cooled to room temperature under the same atmosphere with a 20 °C/min cooling rate. The surfaces of $Ca_{0.5}TaO_3$ films were characterized by an atomic force microscope (AFM), and the root-mean-square roughness is ~ 0.26 nm (Fig. S1). To resolve the crystal structure, both X-ray diffractometry (Cu Kα1) and high-resolution transmission electron microscopy (HRTEM) were applied. For the symmetric X-ray measurement, only $Ca_{0.5}TaO_3$ (001) and (002) peaks were observed on LSAT (001), while only a $Ca_{0.5}TaO_3$ (111) peak was observed when grown on LSAT (111), as shown in Fig. 1 (b) and Fig. 1 (c). Clear Laue-fringes can be seen in both films, indicating highly ordered atomic structure and flat surface, consistent with the AFM results. Using Brag's law, the lattice constants of $Ca_{0.5}TaO_3$ (001) and (111) are 3.919 Å and 2.260 Å respectively. To reveal the epitaxial relation between the $Ca_{0.5}TaO_3$ and LSAT, we performed azimuthal ϕ-scans selecting $Ca_{0.5}TaO_3$ {111} and LSAT {111} for [001] oriented film, and $Ca_{0.5}TaO_3$ {001} and LSAT {001} for [111] oriented film. As can be seen in Fig. 1 (d) and Fig. 1 (f), the films' peaks align exactly with the substrates' peaks, showing fourfold and threefold symmetries for $Ca_{0.5}TaO_3$/LSAT (001) and $Ca_{0.5}TaO_3$/LSAT (111) respectively. The atomic structure of the $Ca_{0.5}TaO_3$ film and LSAT substrate can be readily seen in HRTEM. The cross-sectional HRTEM specimens were prepared by a focused ion beam lift-out technique. As shown in Fig. 1 (f), a sharp interface is seen, and the $Ca_{0.5}TaO_3$ atoms stack orderly on the LSAT substrate confirming their epitaxial coordination.

To examine the electric transport properties of the $Ca_{0.5}TaO_3$ interfaces, we deposited an extra $LaAlO_3$ layer (~ 10 nm) on the top of $Ca_{0.5}TaO_3$ films. Similar to $LaAlO_3$ grown on $KTaO_3$, the $LaAlO_3$ films grown on $Ca_{0.5}TaO_3$ are amorphous, as shown in Fig. 2 (a). A schematic view of the samples is shown in the inset of Fig. 2 (b) where Van-der-Pauw geometry was applied in the measurements. The $LaAlO_3$ films were grown in a vacuum at 650 °C, and the growth details can be found in the Supplemental Materials [22]. As a comparison, $LaAlO_3$ films were also grown on $KTaO_3$ (111) under the same growth recipe. As shown in Fig. S4, a signature of resistance downturn was captured below 2 K, consistent with the superconducting transition temperature reported previously. We present two typical results for $LaAlO_3$/$Ca_{0.5}TaO_3$/LSAT (001) and $LaAlO_3$/$Ca_{0.5}TaO_3$/LSAT (111) in Fig. 2 (b). Different from the $LaAlO_3$/$KTaO_3$, the $Ca_{0.5}TaO_3$ surfaces are semiconducting for both (001) and (111) planes. The room temperature resistances can be as high as ~ $2.5 \times 10^5$ $\Omega/\square$ for the $Ca_{0.5}TaO_3$ (001) plane and ~ $9.3 \times 10^4$ $\Omega/\square$ for the $Ca_{0.5}TaO_3$ (111) plane, whereas the resistances below 50 K are out of our measurement range.

At the $SrTiO_3$ interfaces, the two-dimensional electron gas may originate from polar discontinuity, atomic diffusion, and oxygen vacancies [23-29]. However, the $LaAlO_3$ capping layer in $LaAlO_3/KTaO_3$ turns out to be amorphous, and the polar discontinuity scenario does not apply [7-9]. Local chemistry near the $KTaO_3$ interface has also been examined by energy-dispersive x-ray spectroscopy (EDS) under scanning transmission electron microscopy (STEM) where the atomic intermixing is confined [7]. The carrier densities at the $KTaO_3$ interface are mainly controlled by introducing oxygen vacancies during the growth. By tuning oxygen partial pressure from $2 \times 10^{-8}$ torr to $2 \times 10^{-10}$ torr, the sheet carrier densities at $EuO/KTaO_3$ (111) vary from ~ $3.5 \times 10^{13}$ $cm^{-2}$ to ~ $1.1 \times 10^{14}$ $cm^{-2}$ [7].

To investigate how the $LaAlO_3/Ca_{0.5}TaO_3$ electron transport behaviors are influenced by oxygen vacancies, we grew the $LaAlO_3$ capping layer at higher temperatures. The growth details can be found in the Supplemental Materials [22]. As shown in Fig. 3 (a) and Fig. 3 (b), $LaAlO_3/Ca_{0.5}TaO_3/LSAT$ (001) and $LaAlO_3/Ca_{0.5}TaO_3/LSAT$ (111) remain semiconducting even for the $LaAlO_3$ grown at 800 °C. The carrier densities were measured in Van-der-Pauw geometry with a perpendicular magnetic field. For all the samples we measured, the Hall signals show linear dispersions and the negative slopes indicate the n-type carriers. The carrier density in the $LaAlO_3/Ca_{0.5}TaO_3/LSAT$ (001) can be increased from $5.6 \times 10^{13}$ $cm^{-2}$ to $2.1 \times 10^{14}$ $cm^{-2}$ (for the $LaAlO_3$ growth at 650 °C and 750 °C respectively), Fig. 3 (d). While for the $LaAlO_3/Ca_{0.5}TaO_3/LSAT$ (111) films, the carrier densities increased from $3.4 \times 10^{14}$ $cm^{-2}$ to $4.6 \times 10^{14}$ $cm^{-2}$ (for the $LaAlO_3$ growth at 650 °C and 800 °C respectively), Fig. 3 (e). Different from $SrTiO_3$ whose carrier density can be increased to ~ $1 \times 10^{17}$ $cm^{-2}$ [30], the carrier tunability of the $KTaO_3$ is rather limited (~ $1.1 \times 10^{14}$ $cm^{-2}$) [7]. In $Ca_{0.5}TaO_3$ film, the oxygen vacancies can be introduced during the deposition, right before the $LaAlO_3$ capping layer. Here, a $Ca_{0.5}TaO_{3-x}$ film was grown at 800 °C in a vacuum (growth details can be found in the Supplemental Materials [22]). As shown in Fig. S5, the $Ca_{0.5}TaO_{3-x}$ film peak is much weakened in the XRD measurement, indicating a disordered phase. As expected, the resistance of the $LaAlO_3/Ca_{0.5}TaO_{3-x}/LSAT$ (111) film decreased, Fig. 3 (f). Surprisingly, even the carrier density increased to $9.1 \times 10^{15}$ $cm^{-2}$, the $Ca_{0.5}TaO_{3-x}$ interface kept semiconducting.

Finally, we compared the band structure of $KTaO_3$ and $Ca_{0.5}TaO_3$. We calculated the electronic structure of these two materials utilizing the density functional theory (DFT) within the Vienna Ab initio Simulation Package (VASP) [31, 32], using the projector augmented-wave method [33]. During the calculation, a generalized gradient approximation (GGA) of Perdew-Burke-Ernzerhof type was used to account for the exchange-correlation potential [34]. More calculation details can be found in the Supplemental Materials [22]. As shown in Fig. 4 (a) and Fig. 4 (b), beside the direct band gaps of 1.7431 eV in $Ca_{0.5}TaO_3$ and 2.112 eV in $KTaO_3$ on Gamma points, the projected band structures both show the O 2p orbitals dominated upper valence bands and the Ta 5d orbitals dominated conduction bands. Details on the weights of O 2p orbitals and Ta 5d orbitals for $KTaO_3$ and $Ca_{0.5}TaO_3$ band structures can be found in Fig. S6. In general, $KTaO_3$ and $Ca_{0.5}TaO_3$ are alike in the electronic band structure.

In the $KTaO_3$, the interfacial conductivity can be explained by introducing oxygen vacancies in both the $KTaO_3$ substrate and the $LaAlO_3$ capping layer [35]. Here, even with large amounts of oxygen vacancies introduced into the $Ca_{0.5}TaO_3$, (where the carrier density can be reached to 9.1 x $10^{15}$ $cm^{-2}$, nearly two orders of magnitude higher than the $KTaO_3$ system), the interfaces remain semiconducting. This suggests that the interfacial conductivities induced by oxygen vacancies are disrupted by other mechanisms. As we mentioned above, the polar discontinuity scheme does not apply here since the $LaAlO_3$ films are amorphous. We checked the atomic diffusion near the interfaces of $LaAlO_3/Ca_{0.5}TaO_3/LSAT$. As shown in Fig. 4 (c), elementary mapping was characterized by EDS. Elements like Ca, Ta, La, and Al, are strictly confined to their expected region. However, apart from the LSAT substrate, Sr shows up in the whole layer of $Ca_{0.5}TaO_3$, indicating a significant diffusion. This is different from the $LaAlO_3/KTaO_3$ system, where K and La diffusions are confined, and the atomic intermixing is limited to just a few atomic layers near the interface [7]. We propose that this heavy Sr diffusion disrupts the interfacial metallic conductivity. Furthermore, the absence of superconductivity in $Ca_{0.5}TaO_3$, despite larger carrier densities, highlights the crucial role of specific phonon modes in $KTaO_3$ for mediating superconductivity at its interfaces. For instance, in $KTaO_3$, certain low-frequency phonon modes, such as the ferroelectric soft mode, are suggested to be significant in electron-phonon coupling [11]. The partial occupancy of the Ca site in $Ca_{0.5}TaO_3$ might alter these vibrational dynamics, thus suppressing superconductivity in this material.

In summary, we have synthesized $Ca_{0.5}TaO_3$ films, which have isotopic structure as $KTaO_3$. Structural characterizations indicate that the films are flat and epitaxial to the LSAT substrates. By capping amorphous $LaAlO_3$ layers, $LaAlO_3/Ca_{0.5}TaO_3$/LSAT shows semiconducting behaviors at the interfaces for both (001) and (111) oriented $Ca_{0.5}TaO_3$ films. By further introducing the oxygen vacancies, the interfaces maintain semiconducting, even though the carrier density can be increased to $9.1 \times 10^{15}$ cm$^{-2}$. Detailed elemental characterization by EDS shows heavy atomic diffusion of Sr from LSAT substrate. We propose that besides the oxygen vacancies, reducing atomic intermixing is also necessary for metallic conduction.

This work at Chongqing University is supported by the National Natural Science Foundation of China (Grant No. 12404123, 12374173, and 12374460), the Fundamental Research Funds for the Central Universities (Grant No. 2023CDJXY-0049). We thank Dr. Yan Liu and the Testing Center of Chongqing University for the fruitful discussion, assistance in transport measurement, and structural characterizations.

Fig. 1

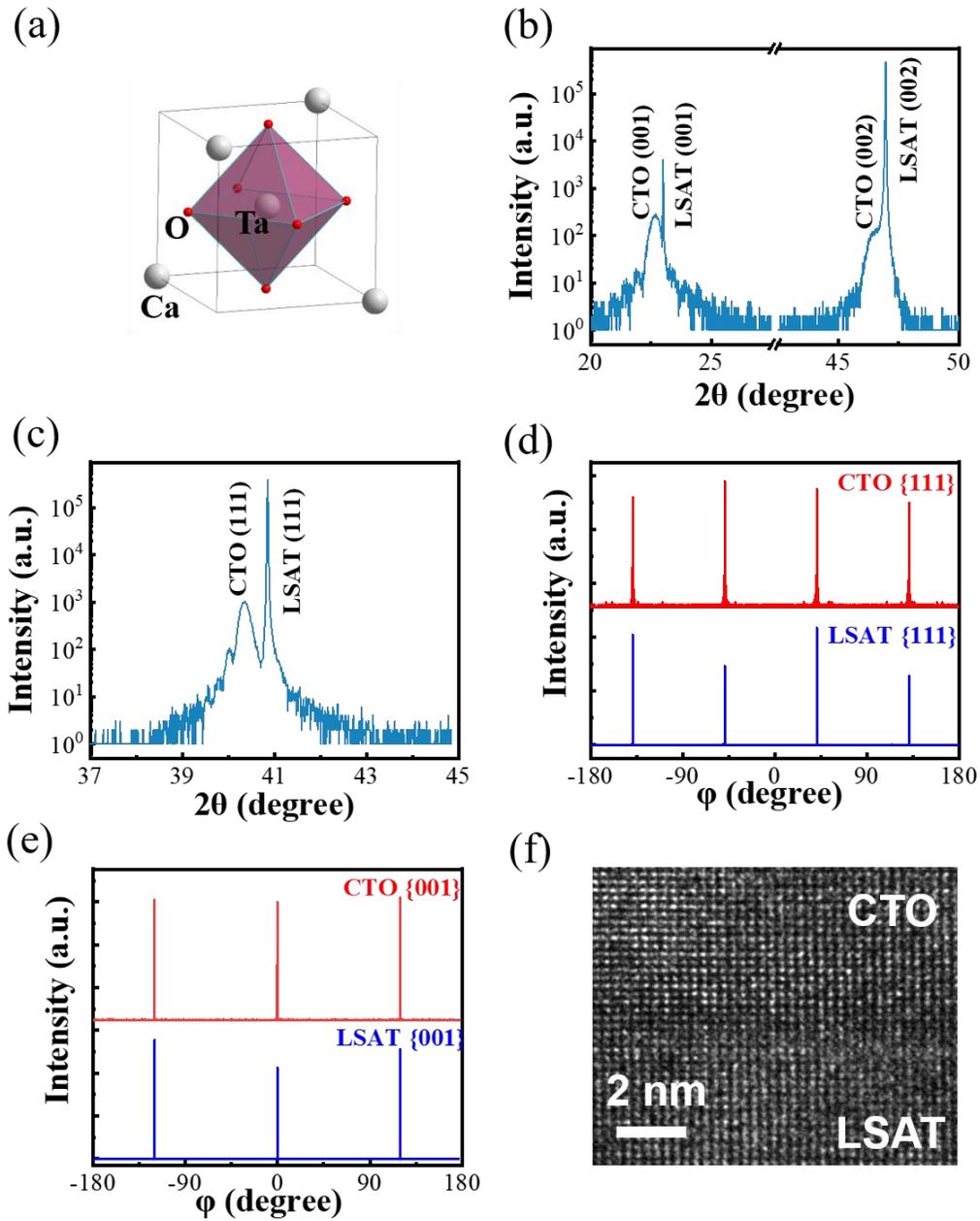

Fig. 1 (a) Crystalline structure of Ca$_{0.5}$TaO$_3$. (b) and (c) X-ray diffraction of Ca$_{0.5}$TaO$_3$ films grown on LSAT (001) and LSAT (111). (d)-(e) Phi scan around CTO {111} & LSAT {111} for CTO/LSAT (001) film and CTO {001} & LSAT {001} for CTO/LSAT (111) film. (f) Cross-sectional HRTEM image viewed along the LSAT [11$\bar{2}$] axis for CTO/LSAT (111) film.

Fig. 2

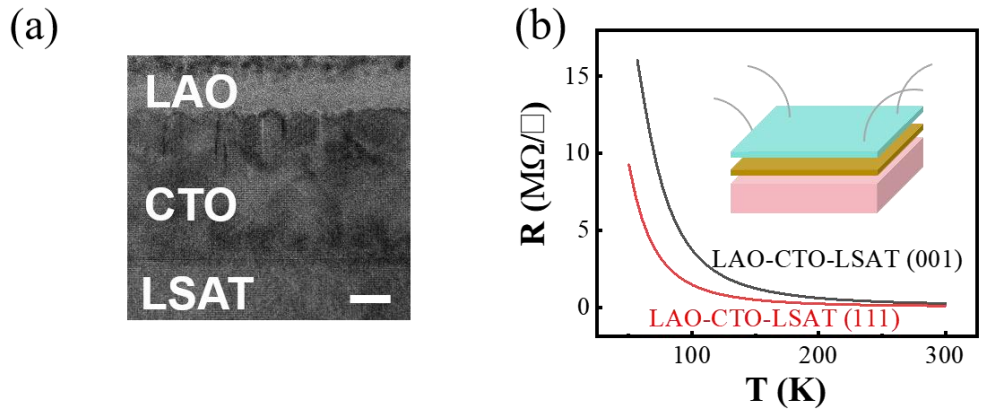

Fig. 2 (a) Cross-sectional HRTEM image showing the amorphous LAO layer. The scale bar ranges 10 nm. (b) Temperature dependences of LAO/CTO/LSAT (001) and LAO/CTO/LSAT (111) films. Inset: schematics of Van der Pauw geometry used in the measurement.

Fig. 3

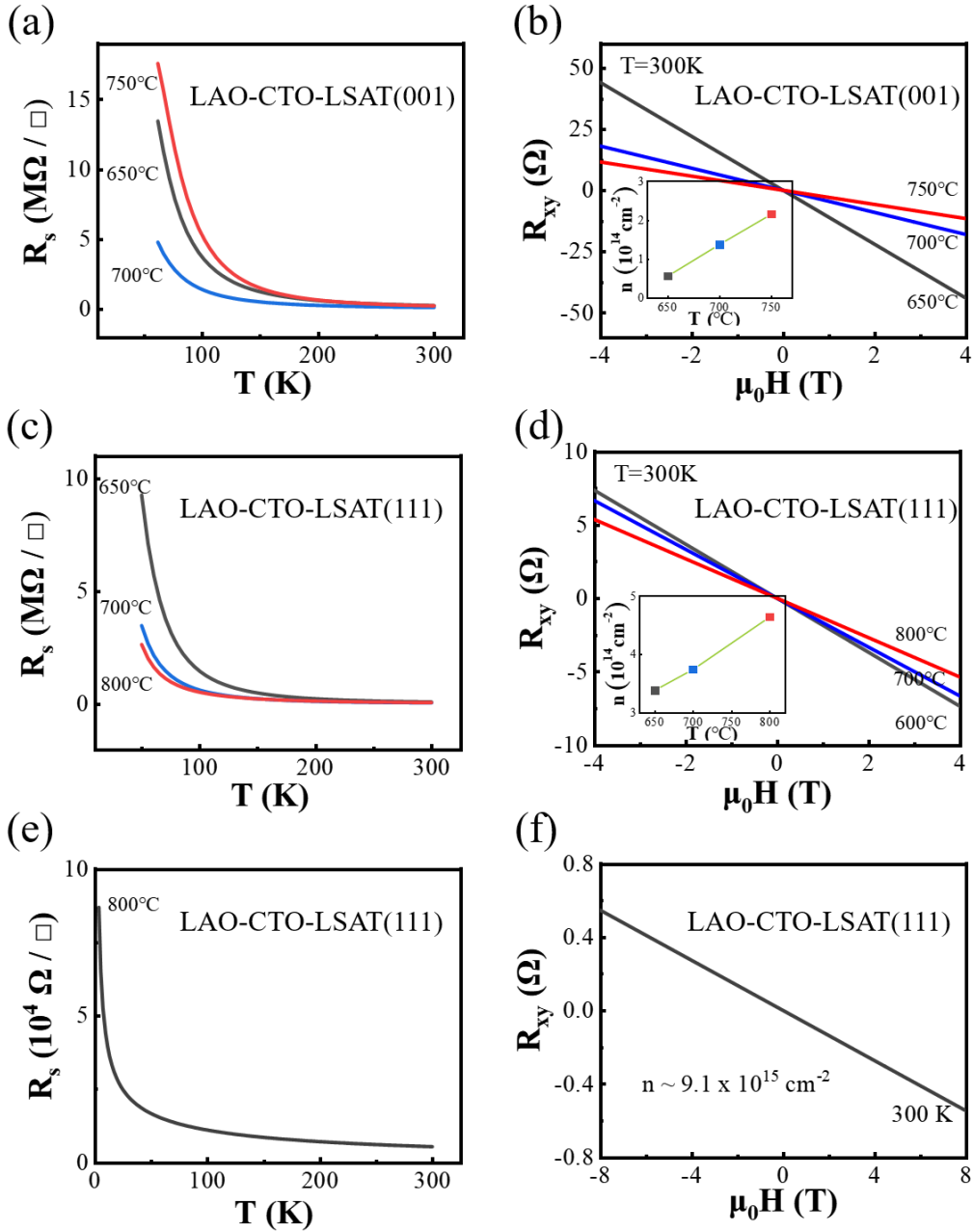

Fig. 3 Temperature dependence of resistance in LAO/CTO/LSAT films. (a) LAO/CTO/LSAT (001) where LAO films were grown at 650 °C, 700 °C and 750 °C. (b) LAO/CTO/LSAT (111) where LAO films were grown at 650 °C, 700 °C and 800 °C. (c) LAO/CTO/LSAT (111) where both CTO and LAO films were grown at 800 °C in vacuum. Hall effect measured at 300 K on the corresponding films from (a), (c) and (e). insets: carrier densities calculated from Hall effect using single band model.

Fig. 4

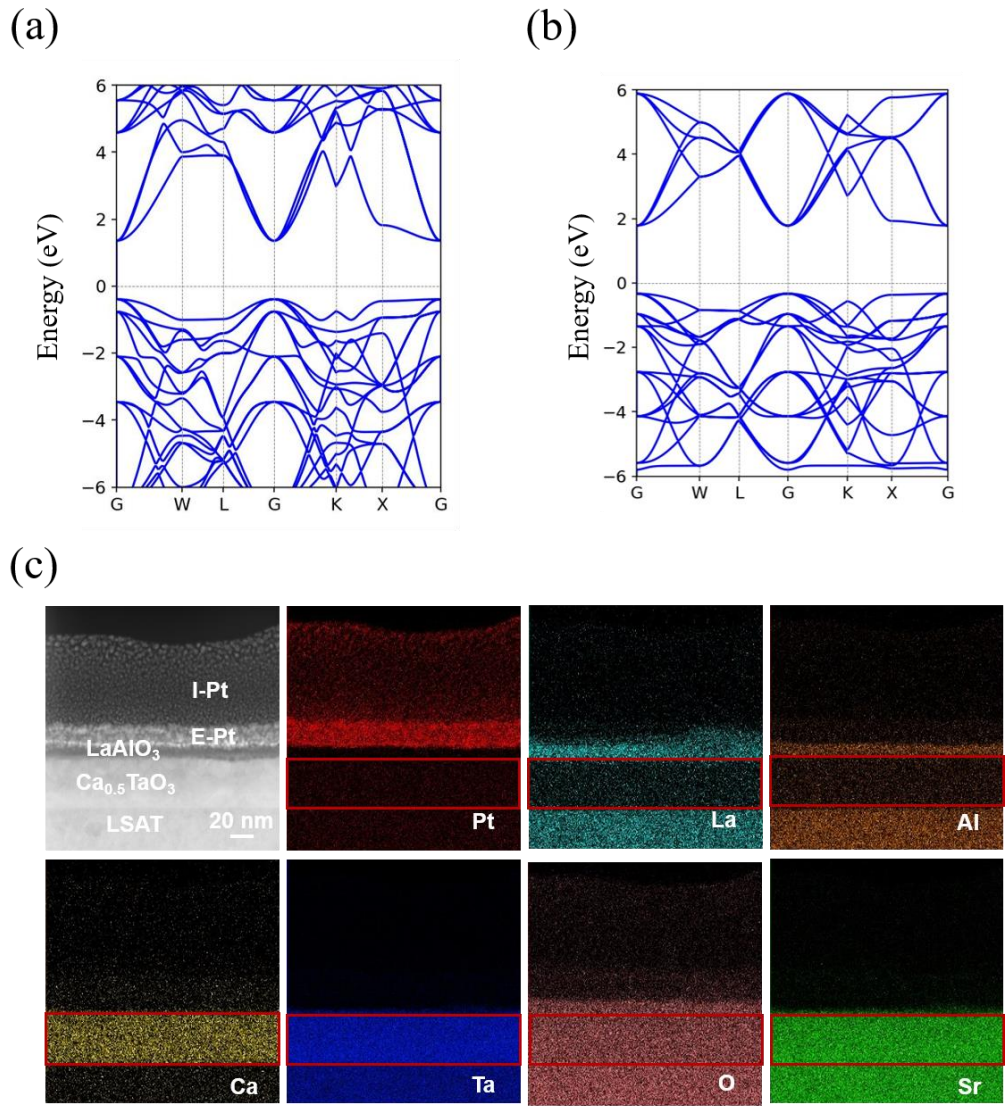

Fig. 4 Band structures calculated from (a) $Ca_{0.5}TaO_3$ and (b) $KTaO_3$. (c) HRTEM image and EDS mapping of the cross section of the LAO/CTO/LSAT (111) sample. Each layer is labeled in the HRTEM image, and the CTO layer is outlined in the EDS mapping.

# Supplemental Material for

## Semiconducting behaviors at epitaxial $Ca_{0.5}TaO_3$ interfaces

Guangdong Nie[1,2,#], Guanghui Han[3,#], Shengpu Huang[4], Huiyin Wu[1,2], Deshun Wang[1,2], Kangxi Liu[1], Hao Ding[1,2], Fangdong Tang[1,2], Licong Peng[3], Dashuai Ma[2,4], Young Sun[1,2], Changjiang Liu[5] & Deshun Hong[1,2*]

[1]Department of Applied Physics, Chongqing University, Chongqing 401331, China

[2]Center of Quantum Materials and Devices, Chongqing University, Chongqing 401331, China

[3]School of Materials Science and Engineering, Peking University, Beijing 100871, China

[4]Institute for Structure and Function, Department of Physics, and Chongqing Key Laboratory for Strongly Coupled Physics, Chongqing University, Chongqing 400044, China

[5]Department of Physics, University at Buffalo, SUNY, Buffalo, NY, 14260, USA

*Correspondence to: dhong@cqu.edu.cn

**This file includes:**
Materials and Methods
Fig. S1 – S7

## Materials and Methods

**Samples.** The samples were synthesized by pulsed laser deposition (PLD). During the growth, a sintered ceramic $Ca_{0.5}TaO_3$ target was used. For the LAO capping layer, a single crystal LAO target (MTI corporation) was used. The substrates were LSAT (001) & (111) and KTO (111) single crystals (MTI corporation). KrF excimer laser with a 248 nm wavelength was applied, and the fluence for depositing CTO and LAO was ~ 2 $Jcm^{-2}$. For samples measured in Fig. 1, Fig. 2, Fig. 3 (a) - Fig. 3 (d), and Fig. 4, the CTO films were grown at 800 °C in 1.0 Pa $O_2$. For samples measured in Fig. 2 and Fig. 4, the LAO layers were grown at 650 °C in vacuum. For samples measured in Fig. 3 (a) – Fig. 3 (d), we varied the LAO growth temperature from 650 °C to 800 °C in vacuum. For samples measured in Fig. 3 (e) - Fig. 3 (f), CTO and LAO layers were grown at 800 °C in vacuum.

**Transport measurements.** All the contacts to the interfaces were made by wire bonding with Al wires. The resistance measurements were carried out in PPMS (Quantum Design) using Van-der-Pauw geometry as schemed in the inset of Fig. 2 (b).

**Calculations.** We calculated the electronic structure of $Ca_{0.5}TaO_3$ and $KTaO_3$, starting with the case without vacancy. The lattice structure of $Ca_{0.5}TaO_3$ and $KTaO_3$, hold the Pm-3m symmetry

(space group No.221), with K atoms taking the Wyckoff position of 1a, Ta atoms taking the Wycoff positions of 1d, and O atoms sitting on the 3c position. Here we take a Face-centered cubic primitive cell of a 2*2*2 supercell to calculate the band structure as it holds two times the chemical formula.

To model the vacancy of $Ca_{0.5}TaO_3$, we removed a Ca atom [site (1/2,1/2,1/2)] in the FCC primitive cell. The lattice constant (reduced perovskite structure) we used for $Ca_{0.5}TaO_3$ was 3.93 Å and 4.02 Å for $KTaO_3$. During the calculation, the kinetic-energy cutoff was set to 500 eV, and a 15 × 15 × 15 Γ-centered k-point mesh was used. A criterion of energy difference in electronic self-consistent calculation is set to $10^{-8}$ eV.

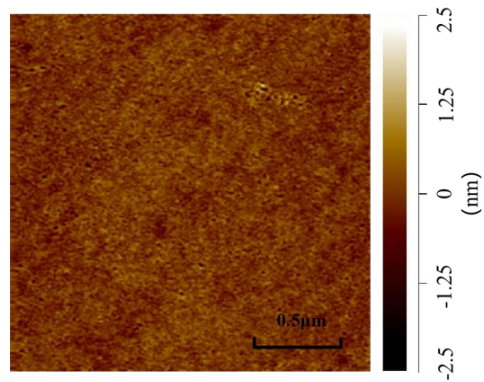

Fig. S1. Surface morphology for a typical CTO/LSAT sample by atomic force microscope. The mean square roughness (2 x 2 um$^2$) is ~ 0.26 nm.

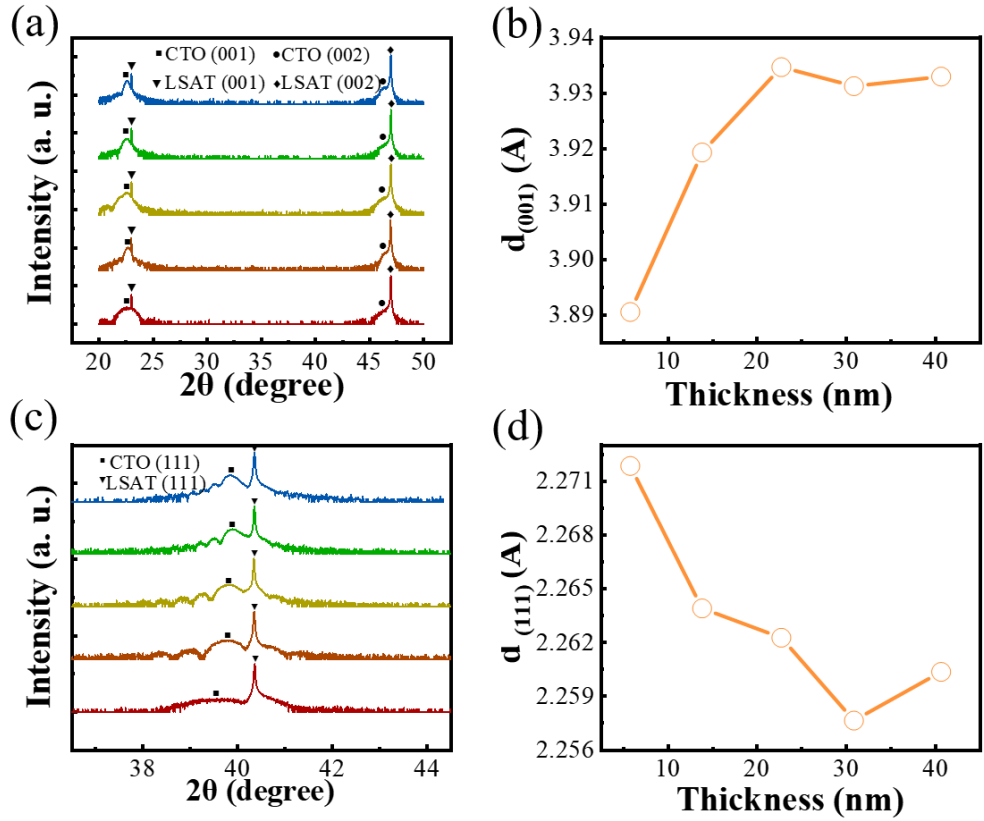

Fig. S2. X-ray diffraction of (a) CTO/LSAT (001) and (b) CTO/LSAT (111) with various thicknesses. (b) and (d) Thickness dependence of lattice constants calculated from (a) and (c) respectively.

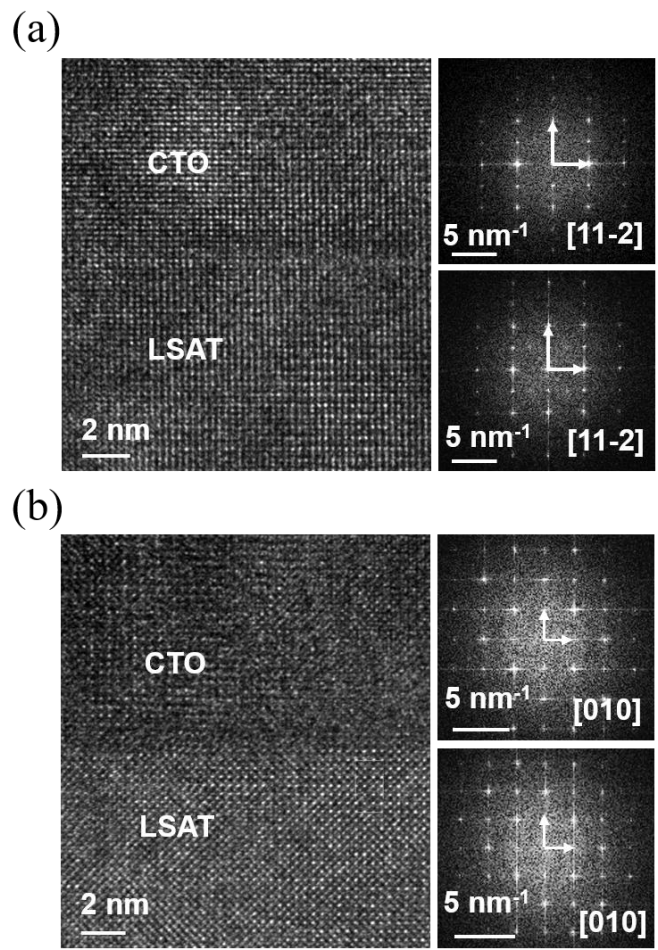

Fig. S3. Cross-sectional images from (a) CTO/LSAT (111) and (b) CTO/LSAT (001). The viewing axes are $[11\bar{2}]$ and [010] respectively. Reciprocal images of CTO and LSAT are shown in the left-up and left-down for each sample.

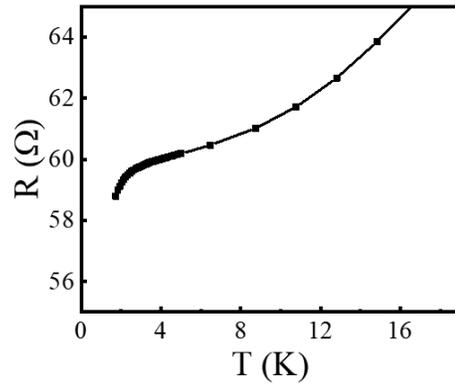

Fig. S4 Temperature dependence of LAO/KTO (111) sample, where the LAO is grown at 650 °C in vacuum.

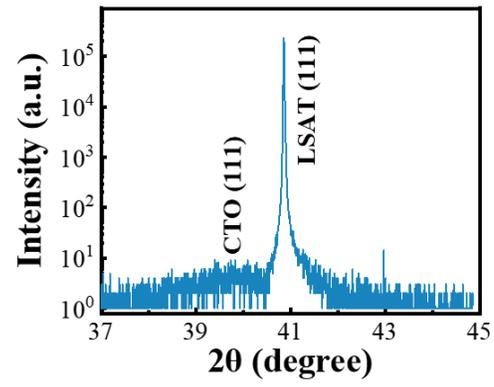

Fig. S5. X-ray diffraction of LAO/CTO/LSAT (111) sample, where both the LAO and CTO layers were grown at 800 °C in vacuum.

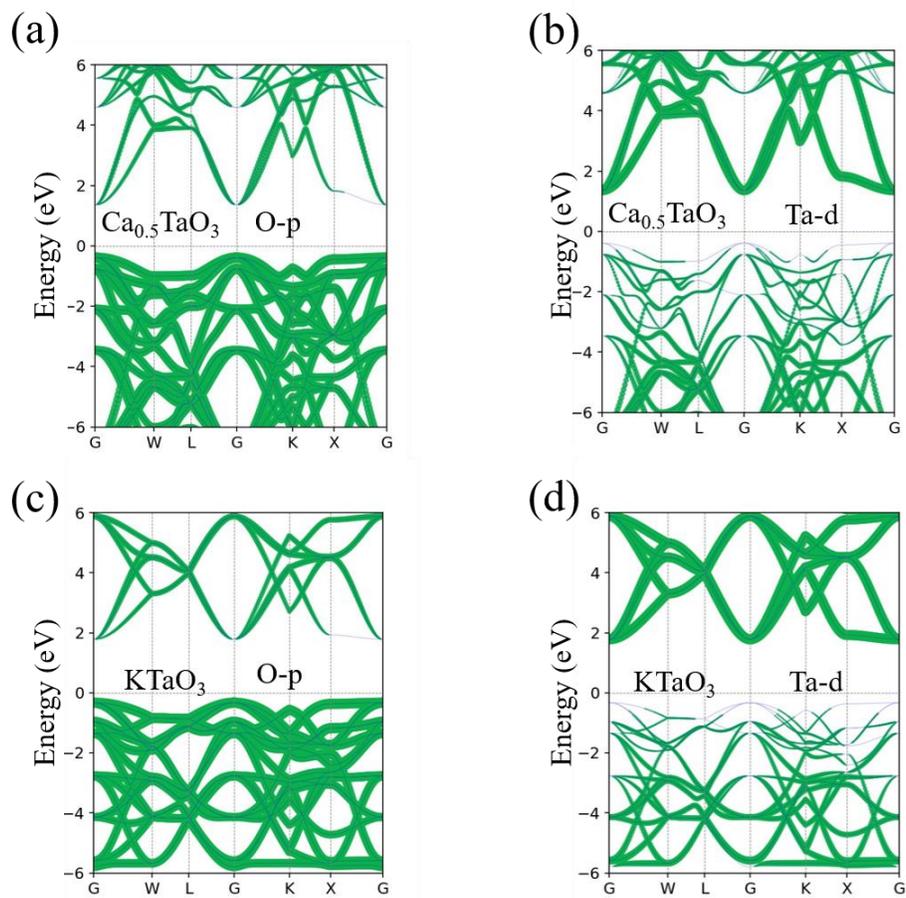

Fig. S6 Band structures calculated from Ca0.5TaO3 (a) - (b) and KTaO3 (c) - (d). Contributions from O-p orbitals and Ta-d orbitals to the band structures are indicated by the line width.

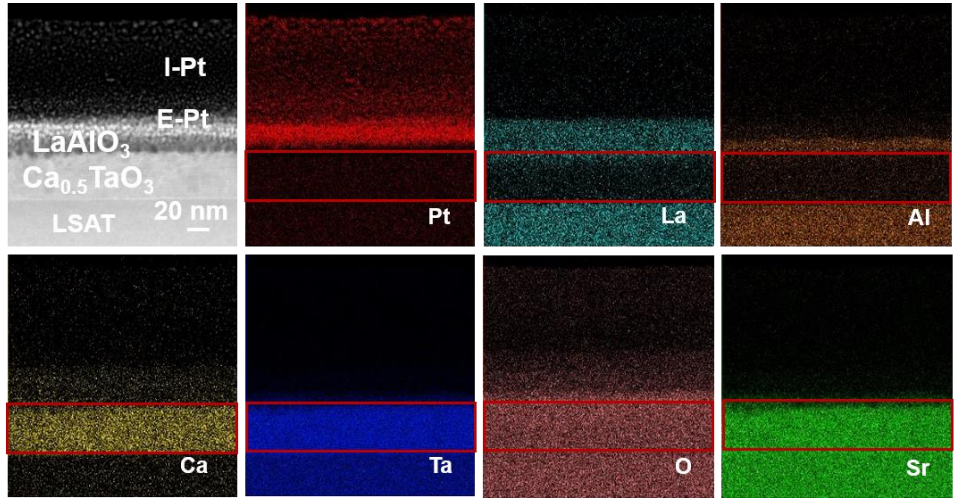

Fig. S7. HRTEM image and EDS mapping of the cross section of the LAO/CTO/LSAT (001) sample. Each layer is labeled in the HRTEM image, and the CTO layer is outlined in the EDS mapping.